# Degradation of Proton Exchange Membrane Water Electrolyzers in Accelerated Stress Tests of Dynamic Load Cycling

Yunyi Zhang[1], Qingbo Gao[1], Jiawei Yang[2,3], Zhen Zeng[1,3], Rui Chen[1,3], Tianyou Wang[1,3], Zhizhao Che[1,2,3,*]

1. State Key Laboratory of Engines, Tianjin University, Tianjin, 300350, China

2. The International Joint Institute of Tianjin University, Fuzhou, Tianjin University, Tianjin 300072, China

3. National Industry-Education Platform of Energy Storage, Tianjin University, Tianjin, 300350, China

*Corresponding author, email: chezhizhao@tju.edu.cn

## Abstract

Proton exchange membrane water electrolysis (PEMWE) is a promising technology for harnessing intermittent renewable energy. This study experimentally investigates the degradation of PEMWE under fluctuating power supply, characterized by dynamic load cycling under accelerated stress test (AST). We focus on the effects of key parameters of dynamic loading, including the peak voltage and cycling frequency, in a set of simplified AST protocols designed to represent selected features of dynamic load fluctuations associated with variable renewable-energy operation. The results unveil the intricate relationship between the structural characteristics of the catalyst layer and the electrochemical performance. An elevated peak voltage accelerates the degradation in the initial phase of the AST. However, a low cycling frequency can mitigate the degradation by limiting the rise in various resistance forms, whereas a higher cycling frequency exacerbates the degradation primarily by increasing mass transport resistance, suggesting a frequency-sensitive deterioration of the system's components.

## 1. Introduction

Proton exchange membrane water electrolysis (PEMWE) is a pioneering technology in the pursuit of a hydrogen economy, powered primarily by renewable energy sources. The attractiveness of PEMWE systems lies in their adaptability, remarkable efficiency, and low environmental impact, positioning them as a key element of our sustainable energy future [1-3]. Renewable energy sources often exhibit an inherently intermittent nature, characterized by rapid fluctuations [4]. For instance, wind energy typically has frequency spikes of 1-2 Hz, while photovoltaic power can experience frequency variations ranging from 0.1 to 0.5 Hz [5,6]. This erratic energy supply underscores the necessity for robust and resilient systems capable of withstanding dynamic operational conditions. Therefore, with the adoption of PEMWE systems, especially in harnessing intermittent renewable sources, it becomes increasingly imperative to understand the PEMWE degradation mechanisms under fluctuating power supply conditions [5,7].

In pursuit of evaluating the durability of PEMWE systems fueled by renewable energy and uncovering the intricacies of their degradation, researchers have developed various accelerated stress test (AST) protocols [8,9]. These protocols were designed to simulate the unpredictable operational conditions and expedite aging processes in a controlled environment. For example, Choi et al. [10] used an AST protocol to condense a day's solar power output into one minute and run it over five days to assess the durability of electrolyzer anodes under different conditions, and found that better-dispersed anodes exhibited lower performance degradation and thus improved durability. Kuhnert et al. [9] designed two AST protocols on PEMWE to replicate real-world solar fluctuations and downtime and found that battery assistance can significantly mitigate component degradation compared to direct solar coupling with frequent load changes. Furthermore, Weiß et al. [11] used AST alternating between different current densities with idle intervals to simulate fluctuating power conditions on a PEMWE, revealing an initial increase in performance followed by a significant drop in performance due to an increase in the high-frequency resistance. These studies aim to optimize PEMWE for grid services, while concurrently evaluating their performance, durability, efficiency, and dynamic behavior within the intricate framework of renewable energy integration [12].

However, the complexity of PEMWE degradation arises from the diversity of conditions and sample compositions across the research protocols, making it challenging to draw consistent conclusions about degradation mechanisms. This underscores the necessity for an in-depth examination of dominant degradation mechanisms under specific AST conditions. Some recent studies have focused on dynamic cycling protocols, introducing variable step durations ranging from minutes to hours [6,13]. These investigations are of particular relevance given the rapid ramp rate events associated with renewable energy sources. Additionally, the impact of high current densities on long-term stability is emerging as a critical research area, bearing potential implications for system cost and performance optimization [14,15].

In a recent study, Voronova et al. [6] introduced dynamic load cycling as part of the AST protocol for PEMWE systems, integrating variable operational conditions with an emphasis on the degradation mechanism under high-frequency cycling and low voltage limits. Similarly, Kuhnert et al. [9] presented an AST protocol to assess the influence of solar photovoltaic energy on electrolyzer operation, highlighting the importance of understanding operational parameters in relation to long-term stability and performance. Fouda-Onana et al. [12] and Rozain et al. [16] investigated the effects of dynamic voltage cycling and current step changes on degradation processes, including membrane thinning, titanium passivation, and catalyst layer degradation. Moreover, the influence of open-circuit voltage (OCV) periods on degradation has sparked debate, with contrasting findings from Weiß et al.

[11] and Rakousky et al. [17], underscoring the intricacies involved in understanding these mechanisms.

In light of the diverse and sometimes conflicting findings in this field, the degradation characteristics of PEMWE systems during load cycling remain unclear. Particularly, the effects of the peak voltage and the cycling frequency, which are two important parameters of dynamic load cycling, remain insufficiently understood. Therefore, we here study the degradation of PEMWEs under dynamic loading in AST conditions. Specifically, our research focuses on the effects of key parameters of dynamic loading, including the peak voltage and cycling frequency. We incorporate a range of peak voltages and cycling frequencies in a set of simplified square-wave AST protocols to isolate the effects of selected dynamic-loading parameters. These protocols are not intended to reproduce the full stochastic nature of real renewable-power profiles, but to provide controlled conditions for identifying degradation trends associated with peak voltage and cycling frequency. By doing so, we seek to provide an in-depth understanding of the degradation mechanisms specific to PEMWE systems, contributing to the advancement of sustainable hydrogen-based energy solutions.

# 2. Experimental Method

## 2.1 Experimental setup

The experimental setup mainly consists of a deionized (DI) water supply unit, a temperature-control unit, a direct-current (DC) power supply, an electrochemical analysis system, and a PEMWE cell, as illustrated in Figure 1a. A DI water supply unit with a peristaltic pump was utilized to control the water feeding rate of the PEMWE cell. The water-feeding pipes and the PEMWE cell were situated within a temperature-control unit to maintain the desired temperature conditions essential for optimal water-splitting operations. Water was delivered at room temperature and was conditioned in the reservoir to match the specified operating temperature before entering the electrolyzer. Two electric heaters were attached to the two sides of the end plates of the electrolyzer. A thermocouple was inserted into the middle of the electrolyzer's anode to monitor the cell temperature. A temperature controller was used to control the operating temperature of the electrolyzer via the electric heaters. A DC power supply (ITECH IT6723C) was wired to the PEMWE cell to control the voltage or current of the electrolyzer operation. An electrochemical workstation (Solartron EnergyLab XM) was connected to the electrolyzer to measure the electrochemical impedance. Electrochemical impedance spectroscopy (EIS) was performed in a two-electrode full-cell configuration, with the cathode connected as the working electrode (WE) and the anode as the counter electrode (CE).

## 2.2 PEMWE assembly

The PEMWE structure is schematically shown in Figure 1b. It has an active area of 4 $cm^2$ (2 cm

× 2 cm). The catalyst-coated membranes (CCMs) had iridium dioxide ($IrO_2$) and platinum (Pt) catalysts with loadings of 2.2 and 1.2 mg $cm^{-2}$ on the anode and cathode, respectively. On the anode side, the porous transport layer (PTL) employed titanium (Ti) mesh accompanied by a microporous layer (MPL), while on the cathode side, the gas diffusion layer (GDL) employed carbon fiber paper (Toray H060). The cathode and anode flow fields were fabricated from high-purity titanium (TA1 grade). The flow fields have parallel channels with a cross-section of 1 mm (depth) × 0.8 mm (width). The components, including the bakelite end plates, flow field plates, polytetrafluoroethylene gaskets, and MEA, were assembled with a uniform torque of 5.5 N·m. In addition, consistent testing parameters were maintained in each set of tests, except for the test parameters of AST, to guarantee the consistency of the tests.

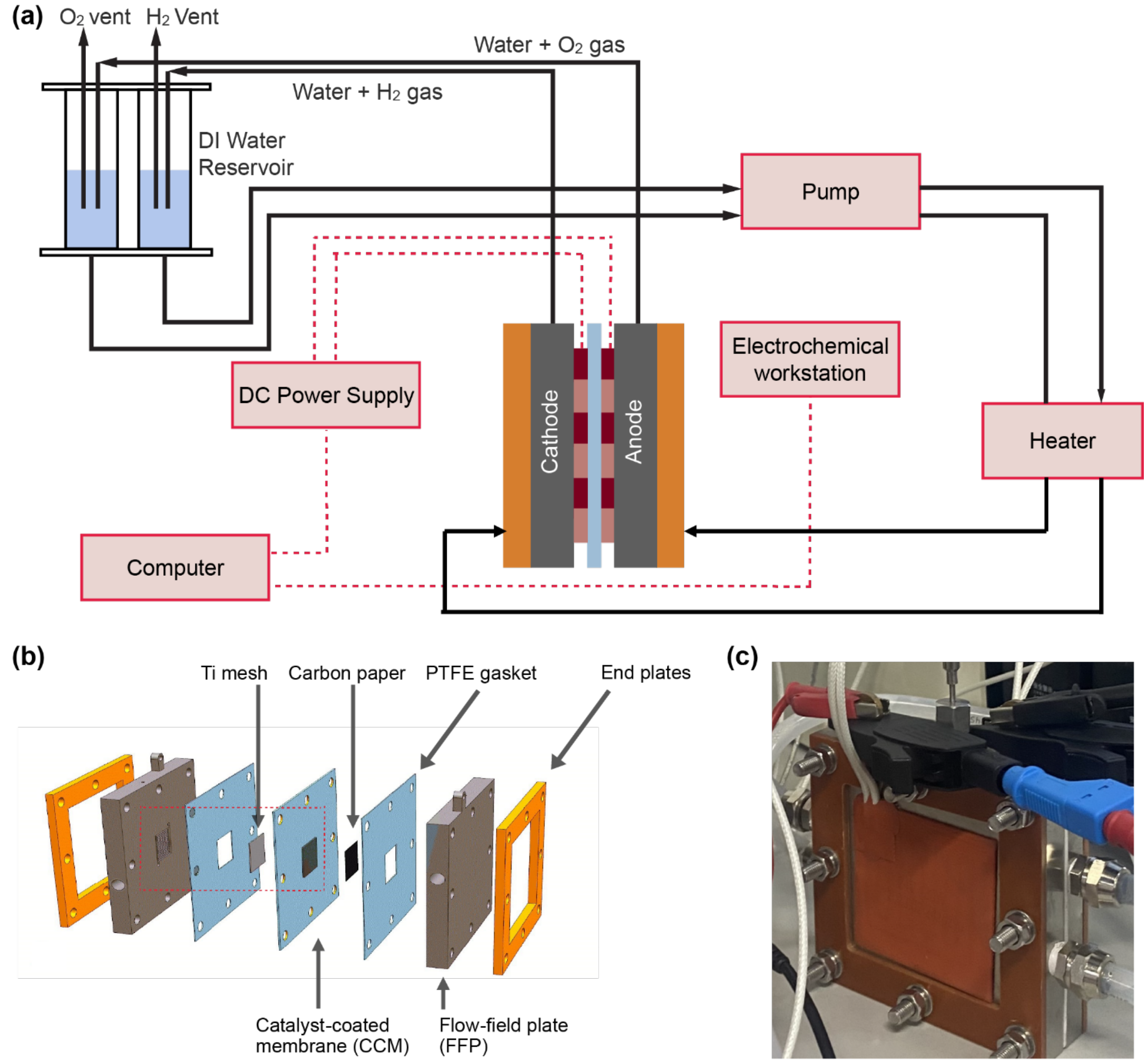

Figure 1. (a) Experimental setup for accelerated stress tests. (b) Schematic of the PEMWE assembly. (c) Image of the PEMWE cell used in the experiment.

## 2.3 Experimental procedure

The PEMWE cell was tested following a rigorous protocol, as illustrated in Figure 2. The tests include three phases, i.e., the Beginning of Test (BOT), the Accelerated Stress Test (AST), and the End of Test (EOT). During the test, the anode and cathode compartments were supplied with DI water at a steady flow rate of 30 mL min$^{-1}$. On the anode side, DI water served as both the reactant and a cooling medium. On the cathode side, although water was not required as a reactant for hydrogen evolution, cathode-side water feeding was used to maintain a comparable thermal and hydration environment across the MEA, facilitate heat removal, and avoid excessive local drying during dynamic operation. The cell was maintained at 80 °C through the dual mechanism of pre-heating the inlet water and employing the electric heaters on the sides of the end plates. Both the anode and cathode outlets were vented to the atmosphere, and no external back pressure was applied. Therefore, all tests were conducted under ambient-pressure conditions.

The BOT phase comprises linear sweep voltammetry (LSV) and electrochemical impedance spectroscopy (EIS). After an activation duration of 30 minutes at a current density of 0.1 A cm$^{-2}$, LSV was measured from 1.3 V to an upper limit voltage corresponding to 1 A cm$^{-2}$. For an all-encompassing electrochemical assessment, EIS measurements were performed in galvanostatic mode across frequencies ranging from 50 kHz to 0.5 Hz at two current densities: 0.3 A cm$^{-2}$, at which the cell operation is predominantly influenced by catalytic kinetics, and 1.0 A cm$^{-2}$, at which the cell operation incorporates more pronounced mass-transport effects. The sinusoidal AC current perturbation amplitude was set to 10% of the corresponding DC current density, i.e., 0.03 A cm$^{-2}$ at 0.3 A cm$^{-2}$ and 0.10 A cm$^{-2}$ at 1.0 A cm$^{-2}$.

In the AST phase, the peak voltage ($E_{\mathrm{peak}}$) and the cycling period ($T$) were used to parameterize the dynamic load cycling, as depicted in Figure 2b. The cycling frequency $f$ is defined as the reciprocal of the period, $f = 1/T$. Therefore, cycling periods of 20, 10, and 5 s correspond to cycling frequencies of 0.05, 0.10, and 0.20 Hz, respectively. To thoroughly assess the impact of $E_{\mathrm{peak}}$ and $T$, the PEMWE cell was evaluated under five AST conditions, as listed in Figure 2a. By referring to the protocol proposed by Aßmann et al. [18], we adopted the potentiodynamic cycling between potentials of 1.4 and 2.0 V with a period of $T = 10$ s and a 50% duty cycle over $10^5$ cycles. After a specific period of time, as shown in Figure 2b, the tests used in the BOT phase (including LSV and EIS) were repeated. It is noteworthy that, for instance in Case 1, the tests were conducted after $5\times10^3$ and $10^4$ cycles (denoted as time points AST1 and AST2, respectively), where the number of cycles was adjusted according to the square wave period to maintain an identical total test duration. In distinct experimental runs, the peak voltage ($E_{\mathrm{peak}}$) was changed to 2.2 V (Case 2) and 1.8 V (Case 3), and the period ($T$) was changed to 20 s (Case 4) and 5 s (Case 5), respectively. To keep the experiment duration at the peak voltage constant, the total number of cycles was changed to $5 \times 10^4$ (Case 4) and $2 \times 10^5$ (Case 5) for different period conditions. In this experimental setting, Cases 1, 2, and 3 are designated for the comparative analysis of the parameter $E_{\mathrm{peak}}$, and Cases 1, 4, and 5 are specifically aligned for the comparative evaluation of the parameter $T$. To reduce the difference between CCM samples used in different cases, each sample was obtained by cutting a large sheet of CCM ($10 \times 10$ cm$^2$). Meanwhile, a voltage-current consistency test was performed before the start of each set of experiments to ensure that the standard performance of 1.54 V was achieved at 0.25 A cm$^{-2}$.

In the EOT phase, the experimental procedure is the same as that in the BOT phase.

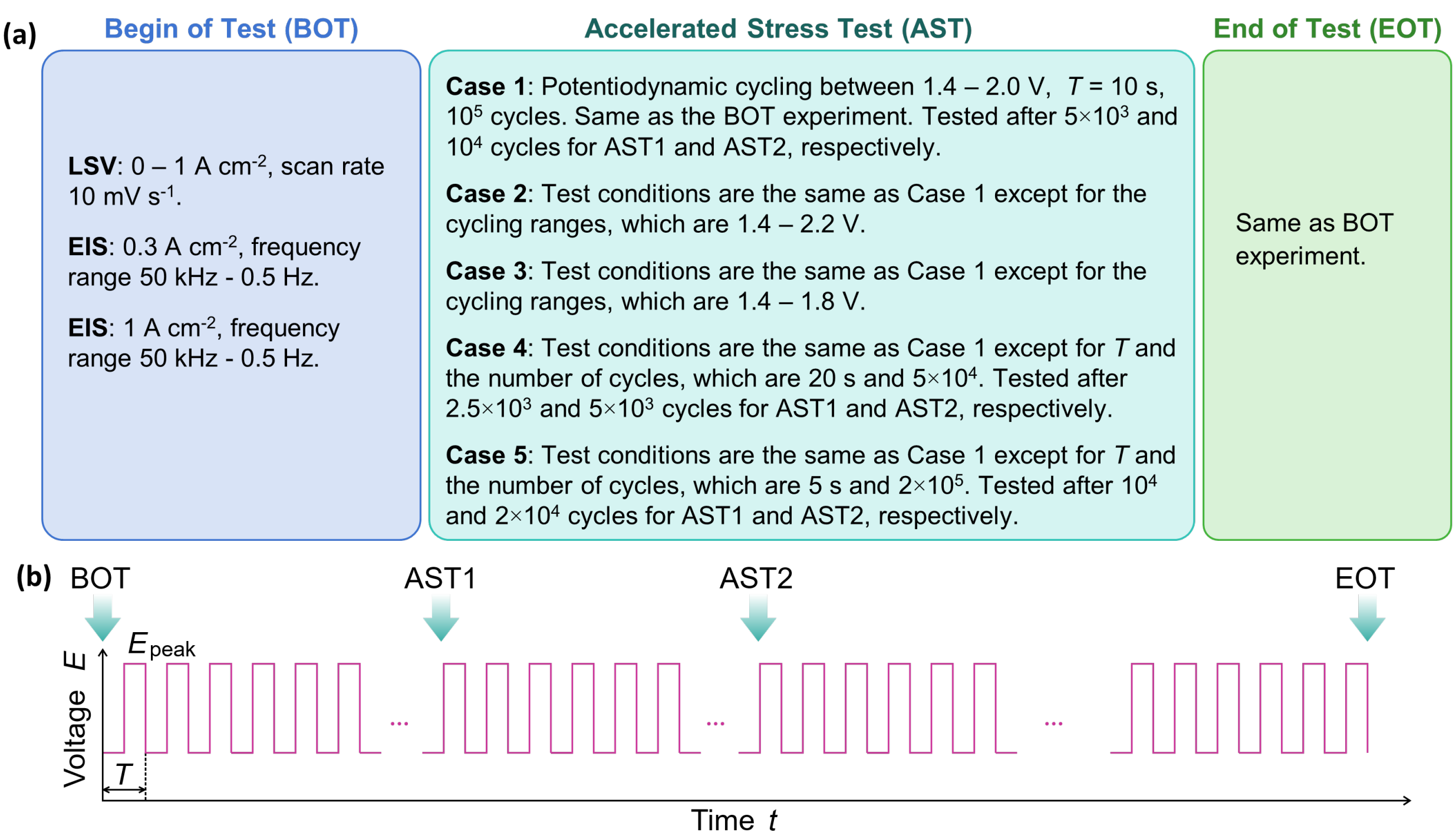

Figure 2. (a) Experimental sequence with different AST conditions (Cases 1-5). (b) Schematic drawing of the potential profile of the dynamic loading cycle in each case.

# 3. Results and discussion

## 3.1 Degradation in cell performance

The polarization curve is an indispensable tool for evaluating the performance of PEMWE systems. The polarization curve at the BOT acts as a benchmark, representing the system's performance in its most pristine state. By comparing with the polarization curves in the BOT and EOT phases, the degradation of the PEMWE cell after the AST can be evaluated. As shown in Figure 3, we can see that Case 4 has the lowest degradation at the EOT, indicating that the experimental parameters in this instance have resulted in minimal adverse impacts on the electrode. Conversely, Cases 2 and 5 at the EOT demonstrate a discernible divergence from the BOT, indicating substantial performance degradation. In Figure 3, all cases have already undergone significant degradation from BOT to AST1. Elevated voltage leads to accelerated performance deterioration of the electrolyzer.

Higher frequency significantly impacts the performance degradation at high current densities.

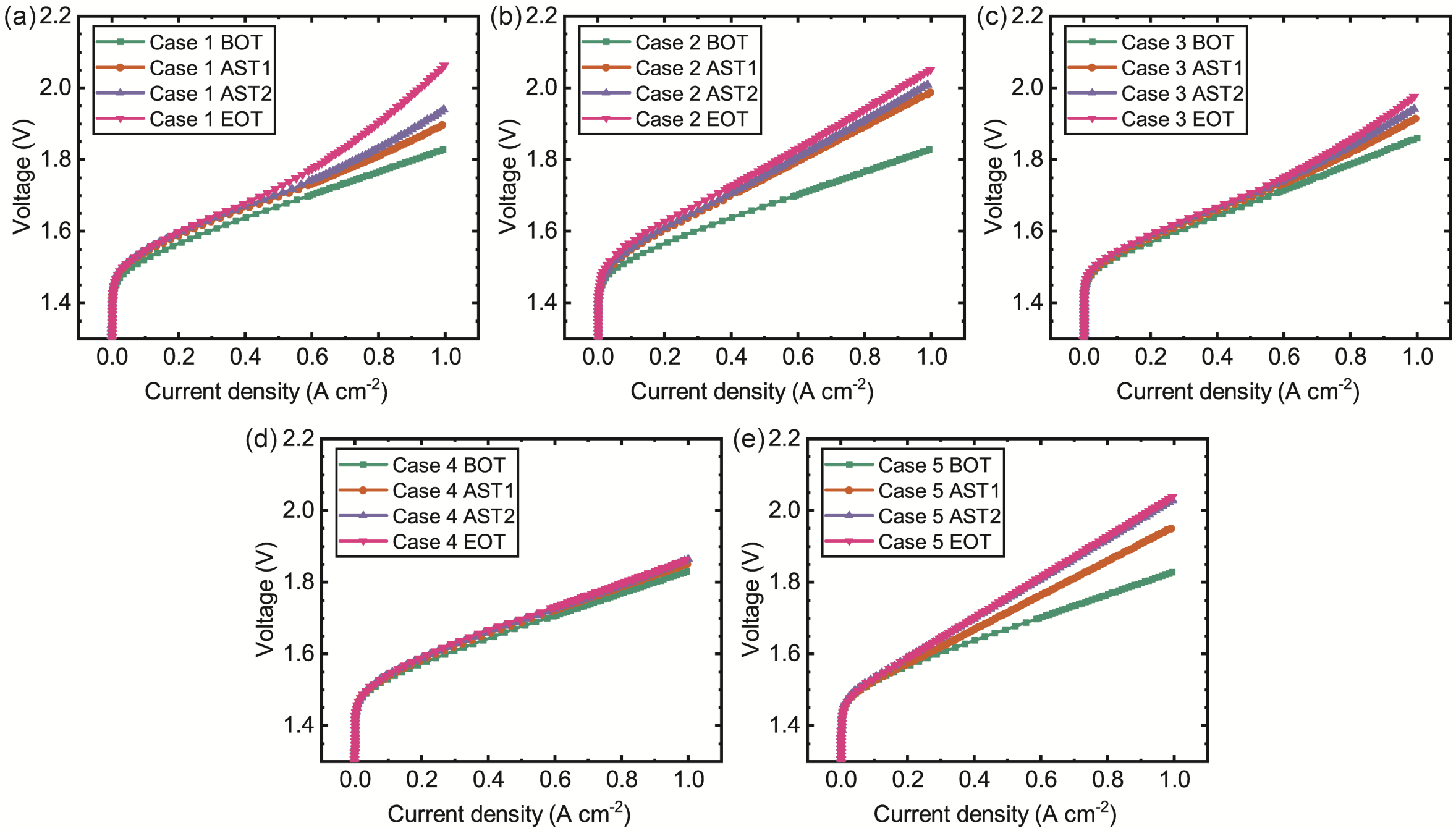

Figure 3. Polarization curves before and after the ASTs under different AST conditions. Panels (a) to (e) show the results for Cases 1 to 5, respectively.

To understand the electrochemical kinetics before, during, and after the AST, we analyzed the Tafel slope based on the linear sweep voltammetry (LSV). The Tafel slopes were obtained by linear fitting of the activation-controlled region of the polarization curves after iR correction for ohmic resistance. The fitting was performed over the current-density range of $0.01 - 0.2\ \mathrm{A\ cm^{-2}}$, where a nearly linear relationship between overpotential and logarithmic current density was observed. The effect of $E_{\mathrm{peak}}$ and $T$ on the Tafel slope can be seen in Figure 4. We can see that the Tafel slopes of Cases 1 and 2 increase sequentially from BOT to AST1, with Case 2 having the largest increase of $17.98\ \mathrm{mV\ dec^{-1}}$. We can also see that a larger $E_{\mathrm{peak}}$ causes greater membrane electrode degradation and leads to serious corrosion of the electrode material during the initial period of the AST. It is related to the fact that a higher voltage led to more intense electrochemical reactions and gas evolution. Moreover, it is worth noting that most of the increase in the apparent Tafel slope occurred from BOT to AST1. Regarding the experimental groups with different cycle periods (Cases 1, 4, and 5), the differences in the increment of the Tafel slope are minimal.

It should be noted that an increased Tafel slope alone cannot unambiguously distinguish between a decrease in accessible electrochemically active area and a change in intrinsic catalytic activity. Therefore, the Tafel slope increase is interpreted here as an indicator of apparent kinetic degradation rather than direct evidence of intrinsic activity loss. The observed kinetic changes may arise from a decrease in accessible active area, changes in the catalyst/ionomer interfacial structure, or variations in intrinsic OER kinetics.

Moreover, the change in the Tafel slope may reflect not only a decrease in the number of accessible active sites, but also possible changes in the apparent reaction pathway or rate-determining step of the OER. However, the present full-cell measurements do not allow the elementary proton-coupled electron-transfer steps to be directly resolved. Dedicated mechanistic experiments, such as H/D isotope-effect measurements under carefully controlled isotope-exchange conditions or operando spectroscopic characterization, would be required to distinguish whether proton transfer, electron transfer, or surface-intermediate formation becomes more limiting after AST [19]. To further elucidate the mechanisms of the degradation under different AST conditions, the changes in the electrochemical performance and the impedance are analyzed in subsequent sections.

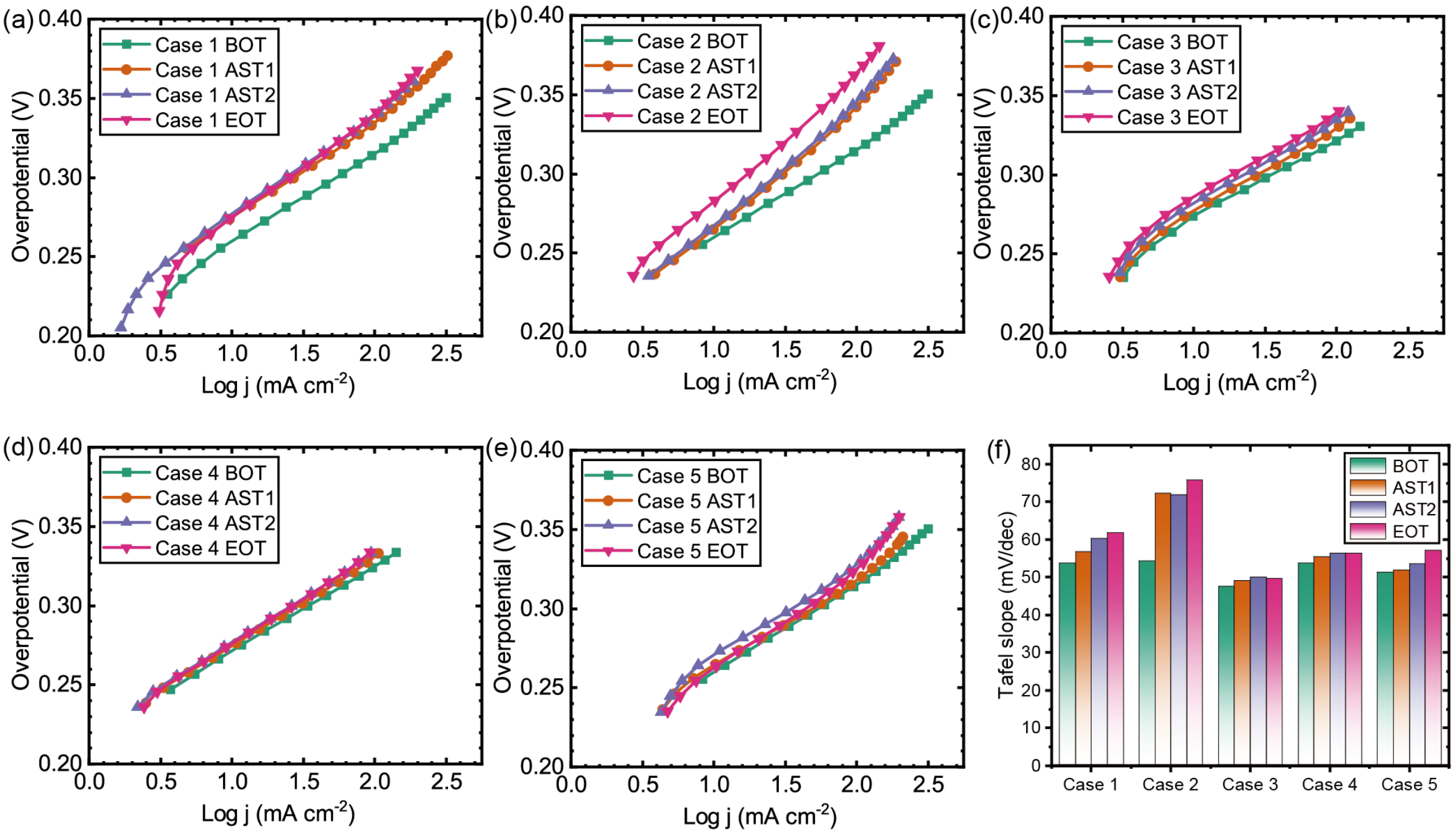

Figure 4. Tafel plots recorded under different AST conditions. Panels (a) to (e) show the results for Cases 1 to 5, respectively. Panel (f) shows the Tafel slope values.

### 3.2 Changes in electrochemical performance

The operational efficiency of PEMWEs is influenced by various overpotentials that arise due to inherent physical and chemical processes. The comprehensive potential governing the water electrolysis in a PEMWE, denoted as $E_{\mathrm{cell}}$, includes several elemental components [20].

$$E_{\mathrm{cell}} = E_{\mathrm{rev}}^{0} + \eta_{\mathrm{ohmic}} + \eta_{\mathrm{kin}} + \eta_{\mathrm{mt}} \quad (1)$$

The reversible cell potential, $E_{\mathrm{rev}}^{0}$, represents the baseline potential difference essential for water electrolysis. Its dependency on temperature can be described by [21]:

$$E_{\mathrm{rev}}^{0} = 1.2291 - 0.0008456 \times (T_{\mathrm{c}} - 298.15) \quad (2)$$

where $T_{\mathrm{c}}$ represents the cell temperature in Kelvin units. The ohmic overpotential, $\eta_{\mathrm{ohmic}}$, is attributed to the inherent resistance encountered during current flow within the cell. It represents the resistances stemming from the membrane, electrode interfaces, and auxiliary contact points. Its quantification can be achieved through EIS, employing the relation given by:

$$\eta_{\mathrm{ohmic}} = j \times \mathrm{HFR} \tag{3}$$

where $j$ is the imposed current density, while HFR is the area-normalized high-frequency resistance obtained from the intercept of the Nyquist plot. Further, the kinetic overpotential, $\eta_{\mathrm{kin}}$, reflects the activation energy required for initiating electrode reactions. Predominantly dictated by the electrochemical kinetics, this overpotential can be deduced using the Tafel relation, given by:

$$\eta_{\mathrm{kin}} = b \times \log10(\frac{j}{j_0}) \tag{4}$$

where $b$ is the experimentally obtained Tafel slope, and $j_0$ represents the exchange current density, indicative of the reaction rate in the absence of any overpotentials. Lastly, the mass-transport overpotential, $\eta_{\mathrm{mt}}$ arises because of diffusional limitations, especially pronounced at elevated current densities. It is influenced by factors such as inadequate reactant supply or impeded product elimination, and can be calculated as follows:

$$\eta_{\mathrm{mt}} = E_{\mathrm{cell}} - E_{\mathrm{rev}}^{0} - \eta_{\mathrm{ohmic}} - \eta_{\mathrm{kin}} \tag{5}$$

Next, the components of the cell voltage were obtained corresponding to the polarization curves, and the breakdown results are shown in Figure 5.

To understand the mechanisms underlying the degradation observed in electrolysis systems, we consider the variation in overpotentials as a function of stress-test cycles, as shown in Figure 5. Notably, Cases 1, 2, and 5 demonstrate a significant rise in the mass-transport overpotential. This increase is consistent with aggravated gas-liquid two-phase effects under high current density, where bubble nucleation, growth, coalescence, and detachment can partially block liquid pathways and reduce the effective active area. As a result, local current density and concentration gradients can become more non-uniform, leading to higher transport-related polarization losses. Repeated bubble growth and detachment may also impose fluctuating local stresses at the electrode/ionomer interface, which can promote gradual microstructural rearrangement and interfacial weakening, thereby reducing the effective electrochemically active area over prolonged cycling [22-24]. The decomposition of the total overpotential into its constituent components reveals that, while Case 2 shows the highest overall overpotential increase, Case 5 exhibits the most substantial increase in the mass-transport overpotentials. These trends indicate that the degradation mechanisms are multifaceted [22,25], potentially involving concurrent processes such as catalyst dissolution [22], metallic cation contamination/poisoning [26,27], and electrode passivation [28]. This interpretation is consistent with the evolution of the EIS after AST (see Section 3.3). In addition, the cross-sectional

SEM evidence of morphological changes supports the possibility of interfacial weakening under cycling (see Section 3.4).

Furthermore, the overpotential decomposition data in Figure 5 also indicate that the performance decay, particularly in Case 1, was predominantly driven by mass transport limitations (a large increment in $\eta_{\mathrm{mt}}$). This aligns with prior studies suggesting that such overpotentials accumulate due to the effects under dynamic AST conditions and high-frequency cycling [29,30]. Conversely, the lower performance degradation in Case 4 (as shown in Figure 3d) can be attributed to a minimal increase in the kinetic and mass transport overpotentials (as shown by $\eta_{\mathrm{kin}}$ and $\eta_{\mathrm{mt}}$ in Figure 5). On the cathode side in Case 1, intermittent operation may induce substantial potential excursions that drive Pt oxidation, dissolution, and redistribution [31], thereby offering a plausible pathway for cathode Pt degradation during load cycling. On the anode side, the observed changes may stem from modifications at the catalyst-ionomer interface that alter the oxidation state of the anode catalyst. Such interfacial/valence-state changes can decrease the affinity for key oxygenated intermediates and hinder water activation, ultimately impairing oxygen evolution reaction (OER) kinetics [32]. Meanwhile, at the longer exposure time in Case 1, the catalyst particles at the anode may grow slightly, leading to significant losses in the mass activity and the surface area in the OER, which results in the degradation worse than in Case 4 [32,33]. For the shortest period $T$ in Case 5, the significant degradation may be due to the extreme cycling frequency, which causes mass transport deterioration.

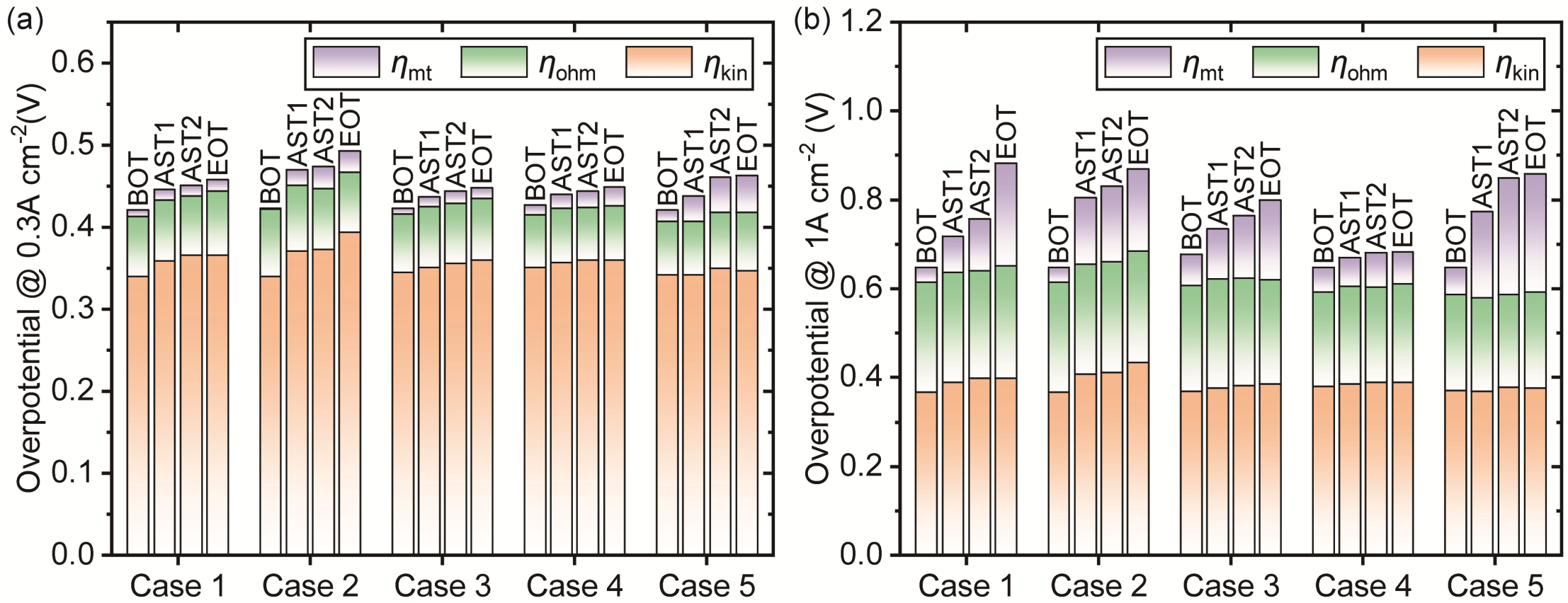

Figure 5. Comparison of the overpotential decomposition before and after the AST at (a) 0.3 A cm$^{-2}$ and (b) 1 A cm$^{-2}$.

### 3.3 Changes in impedance

Electrochemical Impedance Spectroscopy (EIS) serves as an insightful method to discern the impact of operational stress on the CCM's activity. The results extracted from EIS, as depicted in Figure 6, can yield an understanding of the impedance changes in various AST conditions, further elucidating the degradation processes. In addition, an equivalent electrical circuit was used to analyze the EIS data [34,35]. The selected equivalent circuit, $\mathrm{R_{ohm}} + (\mathrm{R_{ct}} \parallel \mathrm{CPE_{ct}}) + (\mathrm{R_{mt}} \parallel \mathrm{CPE_{mt}})$, was adopted based on both the measured Nyquist spectra and commonly used equivalent-circuit interpretations for PEM water electrolysis systems [36,37]. The high-frequency intercept was assigned to $\mathrm{R_{ohm}}$, which mainly includes membrane protonic resistance and electronic/contact resistances. The dominant depressed arc was described by $\mathrm{R_{ct}} \parallel \mathrm{CPE_{ct}}$, representing charge-transfer processes and non-ideal interfacial capacitive behavior. The low-frequency response was described by $\mathrm{R_{mt}} \parallel \mathrm{CPE_{mt}}$, which accounts for mass-transport-related processes, including gas bubble coverage and two-phase reactant/product transport. Therefore, this equivalent circuit provides a physically interpretable description of the ohmic, kinetic, and transport-related losses under the present operating conditions.

To evaluate the reliability of the impedance fitting, the fitting quality was assessed using the ZView-reported Chi-Squared ($\chi^2$) and Sum of Squares values. The fitted spectra showed $\chi^2$ values in the range of 0.000143–0.001873 and Sum of Squares values in the range of 0.010574–0.14796, indicating good agreement between the experimental impedance spectra and the equivalent-circuit fitting.

In evaluating the impedance changes depicted in Figure 6(a-j), we can note the impact of various AST operational conditions on the PEMWE cell. The data reveal distinctive shifts in the ohmic resistance ($R_{\mathrm{ohm}}$), the charge transfer resistance ($R_{\mathrm{ct}}$), and the mass transport resistance ($R_{\mathrm{mt}}$), which are correlated with the electrochemical performance and degradation mechanisms of the cell. The detailed fitting metrics for all EIS spectra are summarized in Table S1 in the Supplementary Material.

In Case 5, which is characterized by a high cycling frequency, the electrolysis cell exhibits a significant increase in mass transport resistance ($R_{\mathrm{mt}}$). A high cycling frequency can accelerate component degradation and exacerbate two-phase transport limitations, manifested as an increase in $R_{\mathrm{mt}}$ at low frequencies due to increased bubble coverage and hindered reactant/product transport [38,39]. Additionally, the observed increase in ohmic resistance ($R_{\mathrm{ohm}}$) reflects multiple underlying effects, such as deterioration of electrical contacts and degradation of the cell's internal components, all exacerbated by the high frequency of operation [40,41].

In Case 3, which is characterized by operating at a lower voltage, there is a slight increase in both charge transfer resistance ($R_{\mathrm{ct}}$) and mass transport resistance ($R_{\mathrm{mt}}$). At the lower peak voltage, the small increase in charge-transfer resistance suggests relatively limited apparent kinetic degradation. The minor increase in $R_{\mathrm{mt}}$ suggests some level of mass transport limitation, possibly due to minor changes in the electrode surface or slight bubble formation, but these effects are less pronounced compared to the cases of high cycling frequencies [38,39].

In Case 2, which is characterized by high voltage operation of the electrolysis cell, a pronounced increase in charge transfer resistance is consistently observed, primarily attributable to the degradation of catalyst materials. In contrast, the variation in ohmic resistance exhibits considerable inconsistency, sometimes increasing and at other times decreasing. This complexity arises from the counteracting effects of membrane thinning, which can reduce resistance, and the potential increase in contact resistance at component interfaces.

Compared with the peak voltage, the cycling frequency has a smaller effect on charge-transfer resistance, while its influence is more evident in mass-transport resistance. As shown in Figure 7, the highest cycling frequency tested produces a pronounced increase in mass transport resistance, whereas the changes in ohmic resistance are comparatively small. By contrast, increasing the peak voltage mainly promotes the growth of charge-transfer resistance, indicating a stronger effect on apparent kinetic degradation.

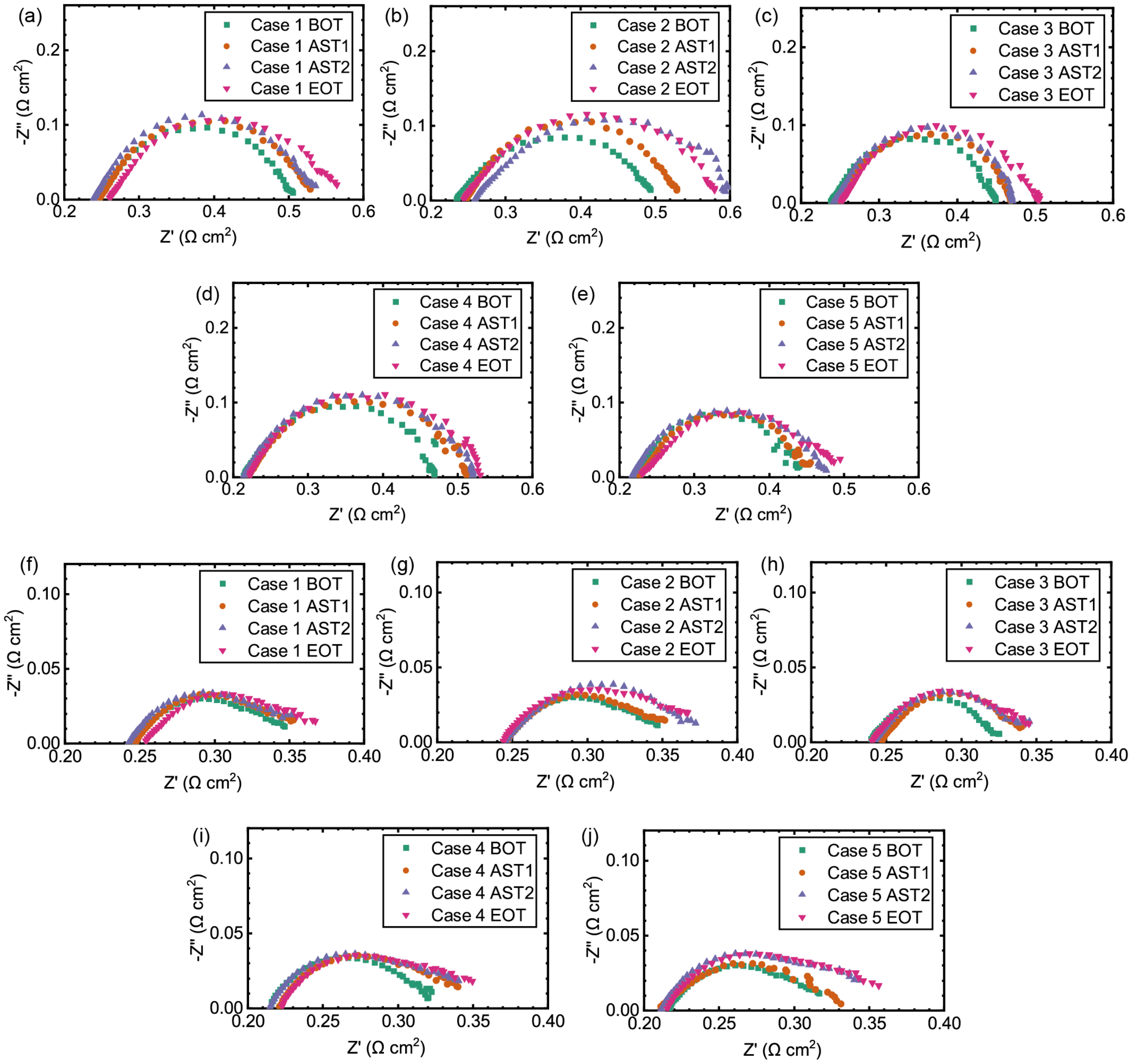

Figure 6. Nyquist plots of the EIS. Panels (a) to (e) show the results at 0.3 A $cm^{-2}$ for Cases 1 to 5, respectively. Panels (f) to (j) show the results at 1 A $cm^{-2}$ for Cases 1 to 5, respectively.

Although the peak voltage and cycling frequency were discussed separately, their potential interaction should also be considered. In the present experimental design, Cases 1–3 isolate the effect of peak voltage at a fixed cycling period, whereas Cases 1, 4, and 5 isolate the effect of cycling period at a fixed peak voltage. Therefore, the current dataset cannot quantitatively determine whether high peak voltage and high cycling frequency lead to additive or synergistic degradation, nor can it evaluate whether a lower cycling frequency can offset the rapid initial degradation induced by a higher peak voltage. Nevertheless, the results suggest that high peak voltage appears to accelerate early apparent kinetic degradation and catalyst-layer morphological evolution, whereas high cycling frequency mainly aggravates mass-transport loss. Their combined effect may therefore involve coupled kinetic, structural, and transport-related degradation pathways, which should be further examined using a factorial AST matrix.

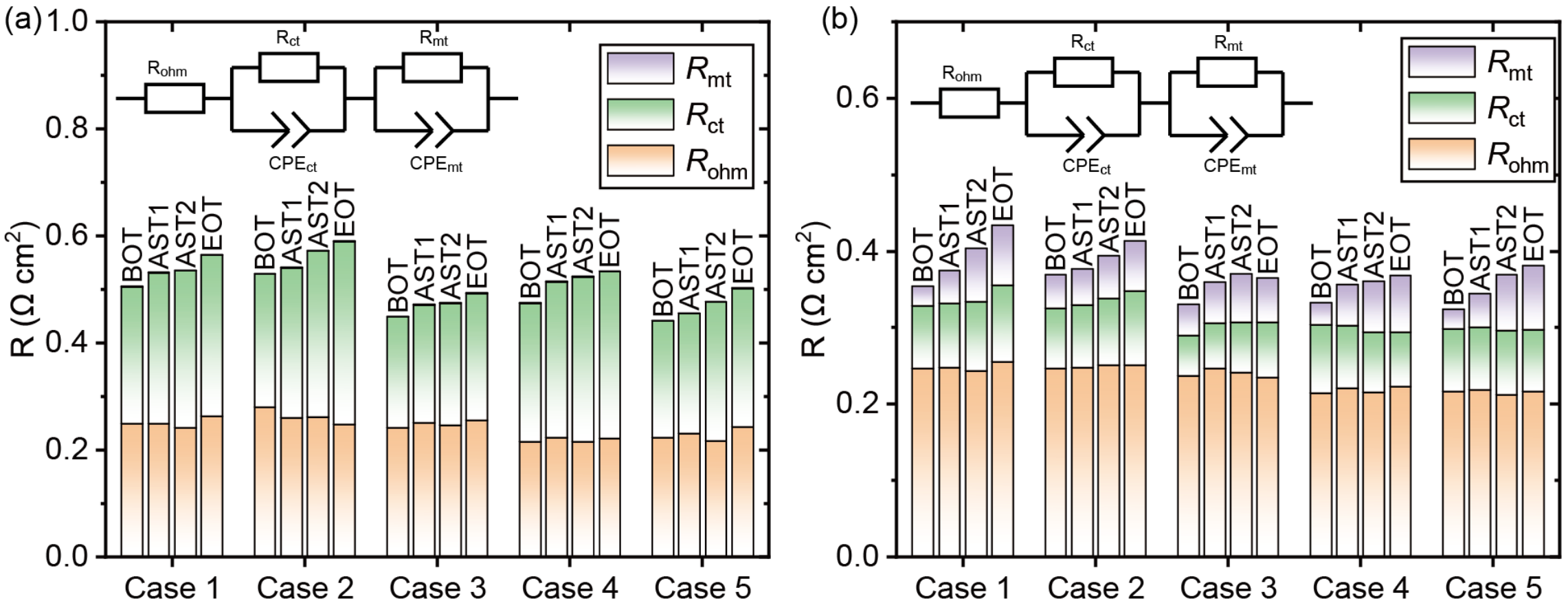

Figure 7. Comparison of the impedance before and after the ASTs at (a) 0.3 A cm$^{-2}$ and (b) 1 A cm$^{-2}$. Insets show the equivalent circuit model used for fitting. Impedance components were obtained from the fitting of the Nyquist plots.

### 3.4 Scanning electron microscopy imaging

Based on post-test characterization, we can observe significant morphological changes in the anode catalyst layer after degradation, as shown in Figure 8. Quantitative thickness measurements were further performed based on the cross-sectional SEM images. The CCM thickness was measured at five different positions in each image. The pristine CCM exhibited a thickness of 168.7±1.3 μm, whereas the thicknesses after AST were 160.5 ± 0.9, 129.4 ± 2.1, 159.3 ± 0.4, 142.6 ± 0.7, and 156.3 ± 1.0 μm for Cases 1-5, respectively. Among these cases, Case 2 showed the most pronounced thickness reduction, which is consistent with the severe morphological degradation observed under the highest peak-voltage condition.

Compared with Cases 1 and 4, the anode side in Case 5 exhibits a more fragmented and discontinuous morphology. This feature may be associated with the high cycling frequency, which repeatedly changes gas generation rates and local bubble coverage within short time intervals. Such rapid gas-evolution cycles can impose fluctuating local mechanical stresses on the catalyst/ionomer network and weaken interfacial contact, thereby promoting fragmentation or partial detachment of the anode catalyst layer. This interpretation is consistent with the pronounced increase in mass-transport resistance observed for Case 5. The thickness variations in the other cases may be associated with local compression, structural rearrangement, and spatial heterogeneity of the degraded CCM. Such severe morphological evolution is linked to the accelerated corrosion and dissolution of catalyst particles under high-voltage/high-current operation in MEA environments [40].

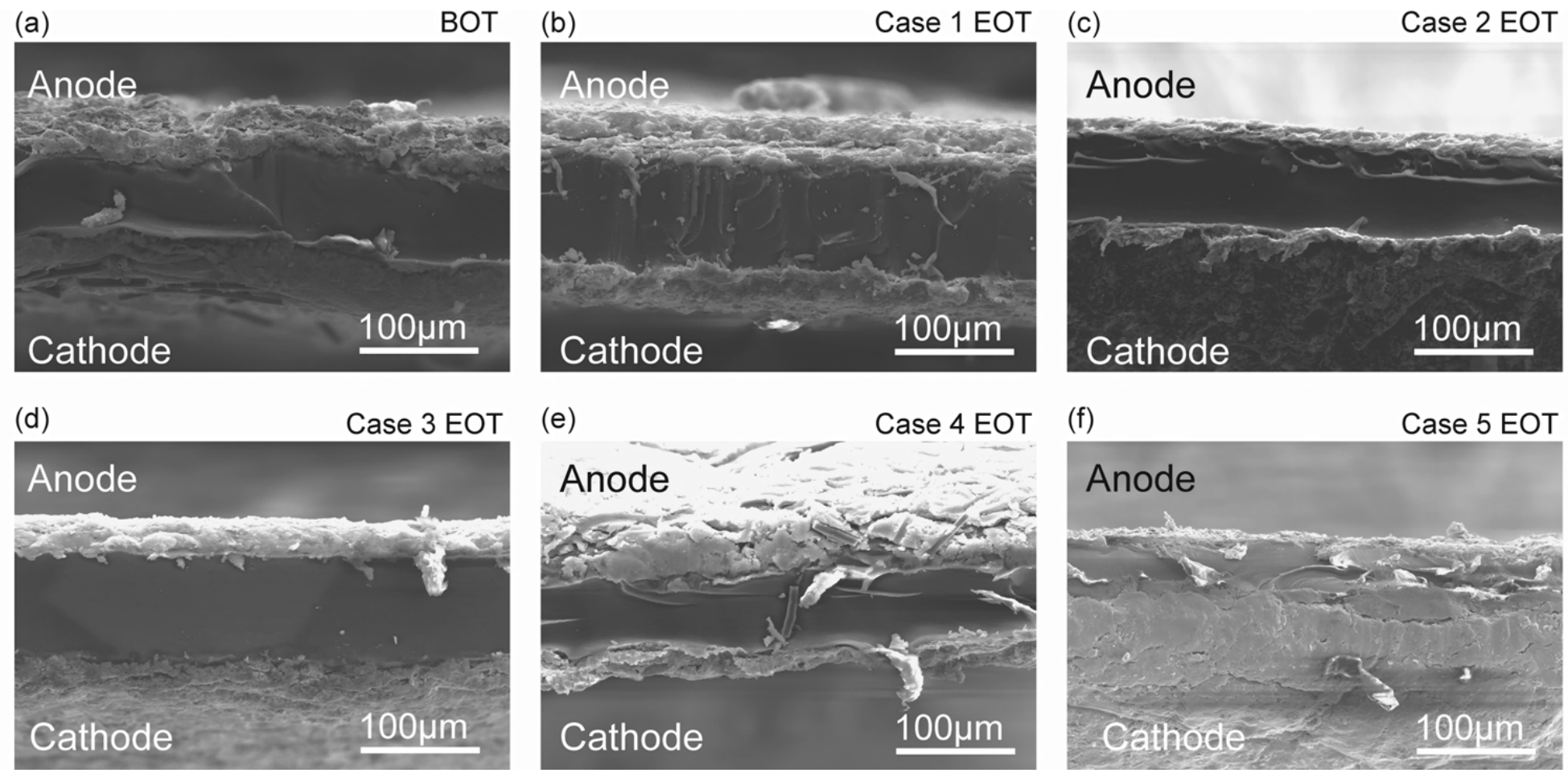

Figure 8. Cross-sectional SEM images of the MEAs: (a) pristine MEA before AST; (b-f) post-test MEAs for Cases 1-5, respectively.

Since the EIS and polarization measurements were conducted in a two-electrode full-cell configuration, the anode and cathode contributions cannot be fully separated. In PEM water electrolysis, the anode OER catalyst is generally regarded as a critical degradation site because it operates under strongly acidic and highly oxidative conditions, where catalyst dissolution, detachment, and catalyst/ionomer interfacial degradation may occur [21,31,39]. Therefore, the observed increase in apparent kinetic polarization is likely dominated by anode-side degradation, as also supported by the pronounced anode morphological changes. Nevertheless, cathode-side degradation cannot be excluded, particularly because Pt oxidation, dissolution, or redistribution may occur under intermittent operation [30]. Moreover, the present SEM analysis mainly focuses on the cross-sectional morphology of the MEA and does not provide detailed cathode Pt particle-size distributions. Thus, the attribution of performance loss to anode degradation should be understood as dominant but not exclusive. Future studies combining electrochemically active surface area (ECSA) measurements, cathode-side particle-size analysis, and operando diagnostics under realistic renewable-power profiles would help further decouple the contributions of anode catalyst degradation, cathode Pt degradation, CCM thickness reduction, and interfacial contact changes.

## 4. Conclusions

In summary, we performed cyclic degradation testing in PEMWE using AST protocols, focusing on the effects of the peak voltage and the cycling frequency. The results show that a higher peak voltage results in faster degradation in the early stage of the AST. A high cycling frequency leads to marked increases in mass transport resistance, indicating accelerated degradation due to operational stress associated with altered bubble growth behavior and subsequent interface deterioration. A relatively lower cycling frequency appears to offer a balanced approach, minimizing increases in different types of resistance and thereby optimizing the lifespan and efficiency of the cell. These results indicate an intricate relationship between the structural characteristics of the catalyst layer and its electrochemical performance under the stringent conditions of AST protocols. It should be noted that the square-wave AST protocol used in this study is a simplified representation of dynamic operation and does not capture the full stochastic fluctuations, ramp-rate distributions, or idle/start-stop events of real renewable-energy-coupled electrolyzers. Future work should therefore validate these degradation trends under realistic wind- or photovoltaic-derived load profiles. Overall, this study provides insight into the degradation mechanisms affecting electrolysis electrodes and highlights the need for robust electrode materials and designs capable of withstanding dynamic operational stresses, thereby supporting the future design and optimization of PEMWE cells with improved efficiency and durability.